\documentclass[runningheads]{llncs}
\usepackage[T1]{fontenc}

\usepackage{graphicx} % Required for inserting images

\usepackage{threeparttable}
\usepackage{float}
\usepackage{hyperref}
\usepackage{tabularx}
\usepackage{graphicx}%
\usepackage{multirow}%
\usepackage{mathrsfs}%
\usepackage[title]{appendix}%
\usepackage{xcolor}%
\usepackage{textcomp}%
\usepackage{manyfoot}%
\usepackage{booktabs}%
\usepackage{algorithm}%
\usepackage{algorithmicx}%
\usepackage{algpseudocode}%
\usepackage{listings}%
\usepackage{csquotes}
\usepackage{multirow}
\usepackage[style=numeric-comp,backend=biber,sorting=none]{biblatex}
\begin{document}

\title{Exploring undercurrents of learning tensions in an LLM-enhanced landscape: A student-centered qualitative perspective on LLM vs Search}

\titlerunning{Higher-ed students' perspective on AI vs Search}


\author{Rahul R. Divekar\inst{1} \and Sophia Guerra\inst{1} \and Lisette Gonzalez\inst{1} \and Natasha Boos\inst{1} \and Yalun Zhou\inst{2}}

\institute{Bentley University\\ \email{rdivekar@bentley.edu, \{sguerra,lgonzalez,nboos\}@falcon.bentley.edu} \and Rensselaer Polytechnic Institute\\ \email{zhouy12@rpi.edu}}

\authorrunning{R. Divekar et al.}
% \renewcommand{\shortauthors}{Divekar et. al.}


\maketitle
\begin{abstract}
Large language models (LLMs) are transforming student learning mirroring how search engines previously disrupted traditional information sources. This study compares these technologies through a within-subjects study where participants learned topics using both Google and ChatGPT, followed by interviews exploring their experiences. Our analysis reveals nuanced insights into students' strategic tool selection processes, showing how contextual factors influence when and why students prefer LLMs over search engines, with particular attention to emergent trust heuristics, epistemic challenges, and evolving notion of authorship and verification--with implications for education stakeholders navigating this technological shift. 

\textit{This version of the contribution has been
accepted for publication, after peer review but is not the Version of
Record and does not reflect post-acceptance improvements, or any corrections. The
Version of Record is available online at: \url{ https://doi.org/10.1007/978-3-031-98465-5_48}. Use of this
Accepted Version is subject to the publisher’s Accepted Manuscript terms of use
\url{https://www.springernature.com/gp/open-research/policies/accepted-manuscript-terms}}
\end{abstract}


% \keywords{Large Language Models, Search Engines, Learning, Higher Education}

% \maketitle

\section{Introduction and Motivation}

\begin{displayquote}
\begin{center}
\begin{minipage}{0.7\textwidth}
\raggedright
\textit{The Homework Machine, oh the Homework Machine,}

\textit{Most perfect contraption that's ever been seen.}

\textit{Just put in your homework, then drop in a dime,}

\textit{Snap on the switch, and in ten seconds' time,}

\textit{Your homework comes out, quick and clean as can be.}

\textit{Here it is— `nine plus four?' and the answer is `three.'}

\textit{Three?}

\textit{Oh me . . .}

\textit{I guess it's not as perfect}

\textit{As I thought it would be. }(\textcite[p.~56--57]{silverstein2009light})  
\end{minipage}
\end{center}
\end{displayquote}

Originally launched as ChatGPT, a large language model (LLM) enhanced with capabilities like search and code interpreters has effectively become a widely accessible “homework machine” for students. It’s free to use, fast, and delivers responses without the obvious errors once imagined by Silverstein in 1981.

A past generation of tools like Google or Wikipedia disrupted the academic landscape by providing efficient access to relevant content. However, the new homework machine contextualizes the information, so it can fit the rubric of a unique homework assignment, masquerading as human written since plagiarism is difficult to detect by both  humans \cite{fleckenstein2024teachers} and machines \cite{wu2023survey}. Educators have responded in various ways--from lunchroom chats to global taskforces--to address how AI tools like ChatGPT should be integrated into education. Yet, there is limited research capturing students’ complex, qualitative experiences with these tools, especially in comparison to established technologies like search. Here, we present qualitative results from an interview with 20 university students who are primarily seen as end-user students rather than experts or researchers of AI. %Of course, any educator can ask students about their AI usage patterns, but imaginably those conversations may not yield  comprehensive or truthful responses from students as students might fear the negative consequences of such disclosure. 

While it’s widely acknowledged that students use ChatGPT for academic work; our study explores the nuances of how, when, and why they do so -- revealing deeper factors such as confidence, evolving trust models, task type, and concerns around reputation. These insights are valuable to educators and higher-education policy makers aiming to understand the student-centered drivers and inhibitors of novel technology in and outside of the classroom. They also offer guidance for technology researchers and developers seeking to take a human-centered approach to designing LLMs for learning. 

\label{sec:intro}

\section{Related Work}
While much existing AIED research focuses on specific applications, commercial LLMs' broader educational impact remains underexplored \cite{ Albadarin2024}. Educators' reactions span from optimism about preparing students for an AI-integrated future \cite{Crcek2023} to concerns regarding academic dishonesty and diminished critical thinking \cite{Kiryakova2023, Mamo2024, Firat2023}. Institutional responses are evident, with 83\% of U.S. universities developing AI guidelines \cite{quinn-2024} and 92\% of provosts reporting faculty requests for AI training \cite{mcdonald2024generative} while some offering broad AI training and access to LLMs \cite{duraisamy-2023}.

Student adoption is widespread, with over 85\% using ChatGPT \cite{Sallam2024}, showing positive attitudes \cite{Chan2023} and diverse applications beyond academics \cite{divekar2024usage, Crawford2024}. This parallels earlier disruptions from Google and Wikipedia \cite{Duha2023, shah2024envisioning}, though LLMs offer unique affordances alongside new risks \cite{Kiryakova2023, Mamo2024}.

However, qualitative research on how students navigate between LLMs and search engines remains limited. This study addresses the gap by investigating how, when, and why higher education students choose LLMs over search engines.

\section{Methodology} 
To examine how LLMs may disrupt traditional learning compared to search engines \cite{shah2024envisioning}, we conducted interviews with 20 university students selected from survey respondents based on completeness and self-rated AI expertise. Student research assistants conducted 19 interviews, with the PI conducting one, to minimize power dynamics. We conducted a within-subjects comparison using two topics (``How does the internet work?'' and ``How does the power grid work?''). Students: (1) wrote about an assigned topic for 10 minutes to establish baseline knowledge, (2) learned about it using either ChatGPT or Google (counterbalanced) for 10 minutes, (3) wrote about the topic post-learning, then repeated with the alternative tool and topic, followed by a semi-structured interview. Key areas explored included: general reflections comparing  learning  with Google and ChatGPT; perceptions of ownership on content produced; confidence, satisfaction, and trust in each tool; overall usage patterns and changes since ChatGPT's introduction; sources of information about ChatGPT; perspectives on privacy implications for both tools; and self-assessed expertise in using these technologies. Participants received compensation, and the study had IRB approval.
The sample predominantly consisted of undergraduates (n=13) aged 18-23 (n=17), with slightly more females (n=11) than males (n=9). Most (n=19) maintained a GPA above 3.0 and majored in business fields including finance (n=6), management (n=5), and business analytics (n=4).
We performed reflexive thematic analysis \cite{Braun_Clarke_2022} on interview transcripts, progressing from initial coding (131 codes) to 11 categories with 57 subcodes, ultimately synthesizing major themes presented in this paper. Analysis involved various tools (NVivo, spreadsheets, whiteboards) with regular team meetings to resolve discrepancies collaboratively. We used LLMs to edit writing in this paper.

\section{Resulting Themes from Qualitative Analysis}

\subsection{Beyond ‘Good’ and ‘Bad’: A Strategic Selection of Tools}

Students adopt a ‘doesn’t hurt to try’ approach with ChatGPT as, compared to Google, as it takes minimal effort to get a coherent response and decide if that is a good result (P10). Students appreciate ChatGPT's continuous interface that unpacks complex information without the interruption of clicking multiple links. P22 explains: ``I want to know how the Internet works and it [Google] [made me go] through three different websites... But for the basics, I want to just understand it in a few simple words.'' ChatGPT offers an informal, judgment-free learning environment where students can request explanations at their level: ``I can say... `how would you explain it to an eighth grader''' (P23) and can take away from classroom engagement: ``some concepts which I might not understand in class, and I might just ChatGPT them [later]'' (P24). Beyond information retrieval, students use ChatGPT as a personalized tutor, generating mock quizzes (P22) or providing step-by-step solutions for math problems (P14). 

Still, students see Google's value for deep research: ``You want to read different kind of articles, get different opinions... when you're going in depth researching'' (P23). Google facilitates incidental learning through exploration: ``...sometimes you come across something that you were not even looking for'' (P11).

While appreciating the immediate benefits of ChatGPT, they recognize potential long-term drawbacks: ``I felt pretty good given I had like a page and a half of notes from the topic I didn't even know at all 10 minutes before... But I don't know... how much I actually understood or I can retain afterwards'' (P21). P6 further explains this tension: ``Instead of absorbing the information, I was more getting it to write it for me.'' Students also express concern about academic reputation when using LLM-generated content: ``I want it to be as accurate as possible because I don't want my professors thinking like `what is this?''' (P22).

Interestingly, trust in ChatGPT varies by topic. P16 explains: ``[for this topic] I feel like the information's been out there for so long. I don't think there's any way I could have pulled false information... `cause it's such a known topic.''

\subsection{A Paradigm Shift From Discerning Sources to Discerning Output and AI Literacy}

While digital literacy programs have addressed how to use search engines, there is a vacuum of formal information sources when it comes to navigating AI as most students indicate learning from informal sources like coworkers (P1, P22, P6); friends, classmates, and common culture (P11, P12, P13); social media and memes (p16, p5), but rarely parents (p9) or educators (P19, P9).  

Participants expressed  that they find prompt engineering to be time intensive and unfulfilling experience (P21, P3) and that there is uncertainty whether the \textit{right} prompt will ever be achieved (P23). Less acknowledged is the realization that to iteratively improve output, one needs to have enough domain expertise to discern good from bad output as pointed by P9: ``It is not as trustworthy when you're relying only on ChatGPT  for all the information you're getting because you can never tell if it's wrong or not.''

\subsection{Navigating the Socio-Technical Landscape: Privacy, Plagiarism, and the Redefinition of Ownership}
Students display notable apathy toward privacy concerns, stemming largely from perceived helplessness. P24 remarks, ``I'm concerned about any privacy... somebody's tracking you, but what can you do about it?'' while P16 suggests resignation: ``Either with Google or ChatGPT, every company has my data already, so it hardly matters.'' Some find misguided comfort in collective vulnerability, with P11 reasoning, ``If my data is at risk, everybody's data is at risk. So I don't believe that bad things would happen to everybody.''

The concept of ownership undergoes significant redefinition with AI tools. While traditional ownership stems from creation effort, with ChatGPT it shifts toward verification effort. As P18 explains, ``If I was 100\% confident that [ChatGPT] was always correct, I would feel almost no ownership because then it's like copying an article and tweaking words.'' Others adopt different perspectives: P9 claims ownership simply because ``[ChatGPT's response] doesn't have the author or research source'' while P5 feels ownership through exercise of will: ``I put in prompts... So I did have some ownership.''

Citation challenges further complicate this landscape. Students wanting to properly attribute sources find themselves unable to do so with ChatGPT, creating both academic integrity and practical citation dilemmas. P18 notes, ``If I'm going to write a paper, I'd rather not put the risk in using ChatGPT... versus using Google and being able to cite it.''

% In all cases, we saw the notion that generated works can are at least partly owned by the student as students  step into a larger complex legal and ethical conversation that is increasingly relevant as AI technologies advance.

% Not every aspect of this personalized learning experience is positive. \textcite{sharma2024generative} have shown that LLMs can trap users in echo chambers. Students recognize that ChatGPT's coherent narrative limits information diversity and incidental learning. Students acknowledge that ChatGPT's quick answers provide hedonistic value while learning, making them feel productive. However, quotes reveal uncertainty about whether they learned correctly or at all. Effectively, in the \textit{input>process>output} learning model, the \textit{process} gets circumvented when using ChatGPT for notes. P13 stated, "I just don't learn anymore... I'll just use ChatGPT, get it done and get the grade"—signaling the absence of Desirable Difficulties \cite{bjork2011making}, challenges that enhance learning. \textcite{bastani2024generative} demonstrated that for math education, ChatGPT acted as a "crutch" that, when removed, negatively impacted students' skills. Outcomes may depend on the level of cognitive engagement with AI-generated information \cite{gajos2022people}. While educators understand this, students are also keenly aware of learning limitations with ChatGPT. 

\section{Discussion}

\subsection{Improved Usability vs. Learning Effectiveness}

Students expressed frustration with fragmented web search results, while ChatGPT provided a seamless learning experience. However, this convenience risks undermining critical thinking by offering quick answers without deep engagement. Echo chamber effects \cite{sharma2024generative} and overreliance risks \cite{bastani2024generative} reinforce concerns. Students appreciated the ease but questioned actual learning without Desirable Difficulties \cite{bjork2011making}. Future AIED tools could integrate LLMs with knowledge tracing to balance user experience with learning outcomes by fostering the zone of proximal development. Educators are encouraged to assess at higher levels \cite{kirkpatrick2006evaluating} e.g., by measuring impact with execution to distinguish student's contribution from ChatGPT's, and ability to think and verify LLM use and outcome critically.

% \rd{Effectively, in the \textit{input>process>output} learning model, the \textit{process} gets circumvented when using ChatGPT for notes. P13 stated, "I just don't learn anymore... I'll just use ChatGPT, get it done and get the grade"}. 

\subsection{An Accelerating Disruption to Community of Learners}

ChatGPT use shifts learning from collaborative human interaction to solitary AI engagement, threatening the social-emotional aspects of education \cite{mccormick2015social, ensmann2021connections}. Students increasingly craft prompts rather than engage peers or instructors. However, instructor assessment-driven accountability still maintains some human connection, as students worry about both grades and their reputational standing with instructors. AIED should design for collaboration, not isolation.

\subsection{The Knowledge Paradox}

As ChatGPT disrupts traditional information seeking, students pivot from evaluating web \textit{sources} to verifying AI \textit{outputs} based on unknown training data. This transition creates a paradox: effectively evaluating ChatGPT's responses requires precisely the domain expertise students lack when seeking AI assistance. This presents educators with dual opportunities to provide unbiased AI education across disciplines and to position themselves as critical guides in students' AI interactions. The AIED community can support this transition through   educator-focused research and workshops, helping faculty who lack AI expertise adapt their teaching practices for responsible AI use within their domains.

\subsection{Vibe Trusting: Topic Perception Outweighs AI Limitations}
In light of the knowledge paradox, students exhibit ``vibe trusting'' as they develop mental models where trust depends on their subjective assessment of a topic's ``established-ness'' to make up for AI's  limitations. This finding extends research on dimensions of trustworthy AI \cite{kaur2022trustworthy}. Vibe trusting also extends but aligns with recent arguments that trust in chatbots relies less on technical understanding and more on social cues and perceived fluency \cite{Magnus_2025}. Overall it points to how students trust stochastic models based on  intuitive feelings while AI systems still remain fundamentally unreliable, creating new challenges for trustworthy AI and digital literacy research.

% Prior research indicates several dimensions of AI trust e.g., fairness, explainability, etc. . Our findings extend the research as we notice students exhibit ``vibe trusting'' as students develop mental models where trust depends on their subjective assessment of a topic's ``established-ness'' to make up for AI's inherent limitations. This phenomenon reveals how students trust stochastic models based on intuitive feelings about topic objectivity, creating new challenges for trustworthy AI and digital literacy research. Future studies should identify which topics trigger this perception-based trust response.

\subsection{Shifting Sense of Ownership}
We saw that some students get a sense of ownership because they verified LLM output shifting the onus from \textit{creation of work} to \textit{verification of work}. Here we see that the biggest flaw of the homework machine works in its favor by creating a sense of ownership for its users through forced verification and iterative prompting. Depending on the extent of verification and iteration, students' feelings on ownership seem in line with a co-constructivist approach. Whereas, other students are more willing to claim ownership even with minimal engagement with the LLM because there are either no sources to cite or because they see ownership as an exercise of their will. This opens new lines of inquiry for assessment, co-construction in learning, and balanced design for human autonomy.

\subsection{Apathetic to Privacy and Security}
Systematic reviews finds that privacy is important for AI to be considered trustworthy\cite{kaur2022trustworthy}.  However apathy  towards privacy been  documented in the context of  social media applications \cite{hargittai2016can} and now seems to extend to AI tools. Students show little resistance to data risks, buoyed by a misplaced sense of collective safety. We see this stemming primarily from a sense of helplessness; AIED practitioners must make students aware and help them make an informed choice.  %This sense of comfort is misguided as massive data leaks affecting large sets of population continue to dominate news headlines. However, none want to stop using technology or pushback in any meaningful way showing that privacy concerns are  a weak undercurrent in the ethical AI landscape.

\section {Conclusion, Limitations, and Future Work}
We reveal nuances in how students strategically navigate between GenAI and web search for self-directed learning. We identified an accelerating shift toward AI tools that undermines collaborative social-emotional learning. Students employ decision frameworks including ease of use and vibe trusting to develop tool-selection shortcuts, and balance awareness of AI limitations with dependence on its convenience. Notably, the possibility of AI hallucinations create unexpected opportunities for students to develop ownership through verification; leaving a paradox when students cannot identify hallucinations.

For AIED, these findings suggest the need for systematic AI education that transcends prompt engineering to cultivate critical awareness of trust, reliability, and epistemic responsibility. Realizing the potential of AI-enhanced education also requires equipping educators with resources for developing students' critical AI literacy across diverse topics.

Study limitations include sample characteristics and the evolving nature of technology. Future research should longitudinally investigate both perceived and observed learning impacts using diverse methodologies. Designers must create socio-technological solutions that balance convenience, trust, ownership, and community-based learning in this new era of \textit{homework machines.}

\section{Acknowledgments}
The authors thank Valente Center for Arts and Sciences and the Faculty Affairs Committee for funding the project. 

\newpage

\printbibliography

\end{document}